\documentclass[twocolumn,showpacs,preprintnumbers,amsmath,amssymb]{revtex4}
\usepackage{graphicx}% Include figure files 
\usepackage{dcolumn}% Align table columns on decimal point 
\usepackage{bm}% bold math 
\def\bx{{\bf x}}              %
\def\br{{\bf r}}              %
\def\bj{{\bf j}}              %
\newcommand{\eq}[1] { Eq.~(\ref{#1})}
\newcommand{\be}{\begin{equation}} 
\newcommand{\ee}{\end{equation}} 
\newcommand{\bea}{\begin{eqnarray}} 
\newcommand{\eea}{\end{eqnarray}}
\newcommand{\av}[1] { \langle {#1} \rangle }

\begin{document}
%--------------------------------------------------------------------
\title{Hyperballistic superdiffusion and explosive solutions to the non-linear 
diffusion equation}
%--------------------------------------------------------------------
\author{Eirik G.\ Flekk{\o}y${}^{2,4,}$}
\author{Alex Hansen${}^{1,3,}$}
\email{flekkoy@fys.uio.no}
\author{Beatrice Baldelli${}^{2,}$}
\affiliation{
${}^1$PoreLab, Department of Physics, Norwegian University of 
Science and Technology, NO--7491 Trondheim, Norway \\
${}^2$PoreLab, Department of Physics, University of Oslo, 
NO--0316 Oslo, Norway\\
${}^3$Beijing Computational Sciences Research Center, CSRC, 
10 East Xibeiwang Road, Haidian District, Beijing 100193, China\\
${}^4$PoreLab, Department of Chemistry, Norwegian University of 
Science and Technology, NO--7491 Trondheim, Norway}
%--------------------------------------------------------------------
\date{\today {}}
%--------------------------------------------------------------------
\begin{abstract}
By means of a  particle model that includes interactions only  via the local
particle concentration, we show  that hyperballistic diffusion may
result. This is done by findng  the exact
solution of the corresponding non-linear diffusion equation, as well
as by particle simulations. The connection between these levels of
description is provided by the Fokker-Planck equation describing the
particle dynamics.
\end{abstract}
%--------------------------------------------------------------------
\maketitle
%--------------------------------------------------------------------

Superdiffusion is characterized by the fact that the root mean
square displacement of some kind of particles,  increases with time
$t$ as  $r_{rms} \sim t^{\tau }$ with the exponent $\tau >1/2$, the 
normal  diffusion value being $\tau=1/2$.
This behavior may arise in physical, biological  or
geological systems;  examples include  Levy flights \cite{bouchaud90,gosh16},  particle 
motion in random potentials or the seemingly random paths of objects
moving in turbulent flows\cite{richardson26,schlesinger87}. 

Biological examples may be found in  the 
foraging movement of spider monkeys \cite{ramos04}   and the
flight paths of albatrosses \cite{vismanathan96,vismanathan99}; in 
both cases  $\tau \approx 0.85$. These movements are Levy walks,
 which  are random walks of  uncorrelated steps of  length $\delta 
\bx$, that take their value from a  distribution $p(\delta x)
\sim 1/\delta x^{\mu +1}$. They result in  superdiffusive behavior with  $r_{rms} \sim 
t^{2/\mu}$ when $0<\mu <2$\cite{bouchaud90}.

However, the mere observation  that the step length
distribution has a fat tail, does not by itself provide any
physical model to explain the superdiffusive behavior.
The simplest physical example  of superdiffusion is perhaps  provided by the  undamped 
Langevin equation which describes a random walk in momentum space and 
a corresponding  real space displacement   with   $\tau = {3/2}$
\cite{jayannavar82}.
This kind of behavior is termed {\it hyperballistic} as $\tau >1$. 
Quantum- or classical particles  in random potentials behave much like
those described by the undamped Langevin equation, and
yield hyperballistic diffusion with  $\tau =3/2$ \cite{jayannavar82} 
too, though  Golubovic et al.  \cite{golubovic91}
studied a case where $\tau = {9/8}$.
 In optical experiments \cite{levi12,sagi12} where 
the spatial coordinate in the direction of the light plays the role of 
the time coordinate, hyperballistic spreading has been observed as well. This effect is linked to Anderson localization \cite{anderson58}, 
and comes from a transition where the light modifies its mean free 
path as it passes through the medium. 

Hyperballistic diffusion seems almost a contradiction in terms, for 
how could a random walker move faster than a directed walker that 
never changes direction? The explanation lies in the fact that the
 velocity, and thus the step length, keeps
increasing with time without limits.  This behavior is of course unphysical in the context of the Langevin equation
as there will always be dissipative forces that match the
fluctuations, but has a physical basis in random potentials.
Without diverging velocities or step lengths, long range
time-correlations are required for superdiffusion, an example being
the {\it elephant  random walk}, so named  because both the walkers
and elephants have long memories,  which in the model give rise to (sub-ballistic) superdiffusion \cite{schutz04}.

Generally, superdiffusion has been modelled by  independent agents
interacting with an environment, or possessing a long term memory \cite{morgado02}.
{\bf The main question of the present letter is if  superdiffusion,
 including the hyperballistic case, could  result directly from a
 Markovian description of particle interactions. }
Such interactive systems could include
crowds of people, bacteria swimmers competing for food \cite{wei98,wu00}
or the evolution of the porosity in a granular packing. For the
purpose of addressing this question we investigate the potentially simplest description of
particle interactions, namely, that  where a conserved concentration $C$ of particles is
governed  by  Ficks  law $\bj = -D(C)  \nabla C$. Here the
$C$-dependence in $D$ reflects interactions between the particles; in many
cases of interest these interactions are well captured by this type of
mean field description. 
Already in 1959  Pattle  \cite{pattle59}  solved the diffusion equation 
\begin{equation}
\frac{\partial}{\partial t} C(\boldsymbol{r},t)  
=\nabla \cdot \left( D(C)  \nabla 
C(\boldsymbol{r},t) \right) \; ,  
\label{eq01}
\end{equation}        
where $C=C(\boldsymbol{r},t)$ is  
the concentration and $D$ is given by  the power law  
$ D=D_0 \left( {C(\boldsymbol{r},t)}/{C_0} \right)^{-\gamma}$ 
where $C_0$ is a constant reference concentration, $D_0$  is the   
diffusivity at that reference value, and the  exponent $\gamma <0$. 
Pattle found  the root mean square displacement $r_{rms} (t) \sim
t^{\tau}$ with 
\be
\tau =
\frac{1}{2- d\gamma } ,
\label{tau}
\ee
 where $d$ is the dimension.  For negative $\gamma $
this will always lead to sub-diffusion.  We have  recently shown
that in $d=1$ there are exact solutions with  positive $\gamma$ as well
\cite{hansen2020},  which still satisfy \eq{tau}, thus 
yielding superdiffusion with $1/2 < \tau <1$ as $\gamma <1$ always.   In the present 
article we take this result further by deriving the solution for
$C(\br , t) $and $r_{rms}(t)$ for  $\gamma >0 $ in any dimension. When $d\ge 2$  the corresponding exponent 
$\tau $ will then take on any value, including those of the
hyperballistic regime, implying that hyperballistic diffusion is a
higher-dimensional effect. We coin the term 'explosive' for the
corresponding time dependence  of $C(r,t)$  because the decay of an
intially localized $C$-profile is qualitatively faster than normal diffusive, or
even superdiffusive, decay.

To validate the mean field description and provide it with a physical
basis,   we  introduce  a  particle model that is described by
\eq{eq01}. The step lengths in this model  $\sim C^{-\gamma /2}$, and
therefore  correspond to
velocities that diverge as $C \rightarrow 0$. This would correspond to an unlimited
access to thermal energy. However, unlike the Langevin equation where
$\tau =3/2$\cite{jayannavar82}, this model can produce any $\tau$-value.

Following  the same lines as   in \cite{hansen2020}
we rewrite  \eq{eq01} as
\begin{equation}
\label{eq06}
\frac{1-\gamma }{D_0C_0^{\gamma}}
\frac{\partial}{\partial t} C(\boldsymbol{r},t)  
=\nabla^2 C(\boldsymbol{r},t)^{1-\gamma}\; .
\end{equation} 
Hence, we see that we need $\gamma<1$ for the equation to be defined when $C(\boldsymbol{r},t)=0$.  
The initial condition at $t=0$ is a point source pulse containing
$N_p$ particles, $C( \boldsymbol{r} ,
0)= N_p \delta (\boldsymbol{r} )$. This means that  there is no intrinsic length- or time scale in
the problem, and the particle number $N(r,t)$  inside a radius $r$
should satisfy the scale invariance condition $ N(r,t) =
N\left(\lambda r , h(\lambda ) t \right) $
for some $ h  (\lambda )$. 
Differentiating this equation with respect to $r$, using the fact that
$dN(r,t) \propto C(r,t) r^{d-1} dr$ leads to the scaling relation
\be 
C(r,t) = \lambda^d C(\lambda r, h(\lambda )t) .
\ee
We are free to chose $\lambda $ such that $ h( \lambda ) t =1$, which
yields $\lambda $ as a function of $t$, say $\lambda = 1/f(t)$, and
\be
    C(r,t)=\frac{1}{f(t)^d}C\left(\frac{r}{f(t)},1\right)=\frac{1}{f(t)^d}p\left(y
\right)\; ,
\label{jhgi}
\ee
where we have  introduced $p(y)\equiv C(y,1)$ and the reduced variable
$y={r}/{f(t)}$. 
Inserting  \eq{jhgi} in  \eq{eq06}  yields
\be
  \label{eq09}
  \frac{\gamma -1}{D_0C_0^{\gamma} (2-d\gamma )}\frac{df(t)^{2-d\gamma}}{dt}
  =\frac{
    \frac{d}{dy}\left(y^{d-1}\frac{d}{dy}p(y)^{1-\gamma}\right)}{
    \frac{d
    }{dy}  y^d p(y)} =c
  \;,
\ee
for some dimensionless constant $c$, which can be absorbed in the
definition of $f(t)$.
The point-like initial condition, implies $f(0)=0$, and the left hand
side of \eq{eq09} can be easily integrated to give
\begin{equation}
    f(t)=\left(\frac{2-d\gamma}{1-\gamma}D_0C_0^{\gamma}t\right)^{\frac{1}{2-d\gamma}}\;.
\label{kjgyr}
  \end{equation}
  Note that this form immediately gives 
\be 
r^2_{rms} = \frac{ \int dr r^{d+1}
 C(r,t) }{\int dr r^{d-1}  C(r,t)} \sim t^{2\tau} . 
\label{lkjiu}
\ee 
with $\tau$ given by \eq{tau}.

From equation (\ref{eq08a}), we also have an expression for  $p(y)$,
\begin{equation}
    \frac{d}{dy}(y^dp(y))=-\frac{d}{dy}\left(y^{d-1}\frac{d}{dy}p(y)^{1-\gamma}\right)\;,
\end{equation}
which can be integrated to give,
\begin{equation}\label{eq10}
    y p(y) +\frac{d}{dy}p(y)^{1-\gamma}=K\;.
\end{equation}
For Fick's law to be valid throughout the domain, $C(r,t)$, and therefore, $p(y)$, must be 
 differentiable everywhere. 
To avoid a spike at the origin we must have $p'(0)=0$ and also a
finite $p(0)$, which  implies that  $K=0$. So, \eq{eq10} may be
integrated to yield
\begin{equation}\label{eq11}
    p(y)=\left[\frac{\gamma}{2(1-\gamma)}y^2+k\right]^{-\frac{1}{\gamma}}
  \end{equation}
where $k$ is an integration constant. This expression is independent of the dimension $d$.
The value of the constant $k$ can be determined through the
normalization, $\int dV C( \br ,t)=N_p$, which gives
\begin{equation}
  k=\left[N_p\left(\frac{\gamma}{2\pi(1-\gamma )}\right)^{\frac{d}{2}}\frac{\Gamma\left(\frac{1}{\gamma}\right)}{\Gamma\left(\frac{1}{\gamma}-\frac{d}{2}\right)}\right]^{\frac{2\gamma}{d\gamma-2}}\;,
  \label{jkhyre} 
\end{equation}
and yields  the concentration field  by means of  Eqs.~(\ref{jhgi}),
(\ref{kjgyr})  and (\ref{eq11}).
The  mean square displacement is given by
\be
r^2_{rms} =\frac{d\pi^{\frac{d}{2}}k^{\frac{d}{2}+1-\frac{1}{\gamma}} }{2N_p} 
\frac{\Gamma\left(\frac{1}{\gamma}-\frac{d}{2}-1\right)}{\Gamma\left(\frac{1}{\gamma}\right)}
    \left(\frac{2(1- \gamma )}{\gamma}\right)^{\frac{d}{2}+1} f^2(t),
\label{jmhvyd}
\ee
which is limited to the range of $\gamma$-values where the integrals
in \eq{lkjiu} converge.
Since $ r^{d+1} C(r,t) \sim r^{d+1-2/\gamma }$ for large $r$ this
range is  $0<\gamma<{2}/{(d+2)}$.
However, in any particle 
simulation there will always be a largest particle position $r_{max}$
that will act as a cut-off. This  means that the  
 $t^{2\tau}$ factor in $r^2_{rms}$ survives, but that its prefactor
 will fluctuate with the $r_{max}$- value. The behavior with different
 $d$ and $\gamma$ is summarized in table \ref{table}. 

\begin{table}
  \centering 
\begin{tabular}{||l|c|c|c||} 
 \hline 
    &    $\tau > 1$ &    $r_{rms} $ prediction converges &  $C$ is normalizable \\  
 \hline  \hline  
d=1 &     never&     $\gamma < 2/3$  &     All $\gamma < 1$ \\
 \hline   
d=2&    $\gamma >1/2$ &     $\gamma < 1/2$  &     All $\gamma <1$ \\
 \hline   
d=3 &    $\gamma >1/3$&     $\gamma < 2/5$  &        $\gamma <2/3$ \\
 \hline   
\end{tabular}
\caption{Behavior with $\gamma$ in various dimensions $d$ as predicted  
by \eq{tau}.}
\label{table}
\end{table}
% Fig 1 a b eq (5)  
\begin{figure}[h!]
\begin{center}
\includegraphics[width=0.70\columnwidth]{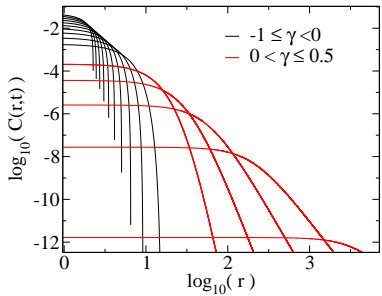}
\caption[]{ The predicted/theoretical concentration field at different  $\gamma$-values when  
  $D_0=1$ and $t=10$.  The 
  black curves show Pattles \cite{pattle59}  solution for  $\gamma = -0.1, -0.2,
  -0.4, -0.5$.
}
\label{jhga}
\end{center}
\end{figure} 
\begin{figure}[h!]
\begin{center}
\includegraphics[width=0.7\columnwidth]{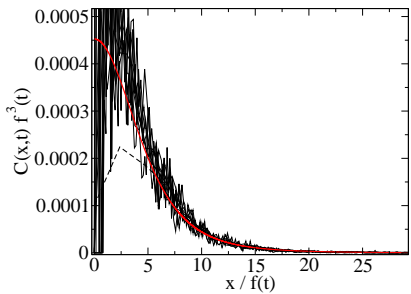}
\caption[]{ Simulations, sampled at equispaced time intervals (the stapled curve shows
  the first time) using  $\gamma =0.4$ and the finite-interaction 
  range model. The curves show $p(y)=C (r,t) f^3(t)$  as a function of $y=r/f(t)$   compared to theory (red curve) of \eq{eq11} and \eq{jkhyre}.  
}
\label{jhgb}
\end{center}
\end{figure} 
We will employ  two simulation models,  both in $d=3$  with  $N_p$ random
walkers, labeled $i$, that have positions  $\br_i \rightarrow \br_i + \delta \br_i$.
The   particles interact only via the  value  
of $C$, which is the local population density. 
The steps  are chosen isotropically at each time step;  to find their
length we 
need to derive the appropriate  Fokker-Planck equation and match it to \eq{eq06}.
For every time step $\Delta t$ the walkers move
\begin{equation}
\delta r_{i\alpha} = \eta g(C(\br_i)) \sqrt{\Delta t} 
\label{kjbyf}
\end{equation}
where $\alpha$ is a Cartesian index and the function $g(C)$ is to be determined.
This defines  a Wiener process with $\eta$ as a random variable with
$\av{\eta}=0$ and  $\av{\eta^2}=1$.
Now, following the  same steps as in  \cite{hansen2020,k07}  we use the
standard Chapman-Kolmogorov, or master equation, to derive the following
Fokker-Planck equation for the particle concentration $C(\br ,t)$
\begin{equation}
\frac{\partial C(\br,t) }{\partial t }
=
\frac{1}{2}\nabla^2 \left( a_2(\br) C(\br,t)  
\right)\; .  
\label{hkuft}
\end{equation} 
Here $a_2(\br)$ is the mean squared jump length per time,
\begin{equation}
a_2(\br) = \int d^3x \frac{\bx^2}{3} W(\br,\bx) =\frac{1}{3}\frac{\av{\delta \br^2}}{\Delta
  t} = g(C)^2\;,
\end{equation}
where $W(\br,\bx)$ is the probability per unit time that a walker
jumps a distance $\bx $ from $\br$.  Setting $g(C) = b C^{-\gamma /2}$ gives
\begin{equation}
\frac{\partial C}{\partial t}
= \frac{ b^2}{2} 
\nabla^2 C^{1-\gamma}\;, 
\label{eq055-2}
\end{equation}
and requiring equivalence with 
\eq{eq06} thus implies that
$
b^2 = {2D_0}/{( C_0^{-\gamma}(1-\gamma ) )} $. This  leads to the step
\be
\delta r_{\alpha} = \eta \sqrt{
  \frac{2D_0 \Delta t  }{(1-\gamma ) 
} } \left( \frac{C (\br ,  t)}{C_0}\right)^{-\gamma /2},
\ee
where the random variable $\eta $ is given above. This defines the particle model that is described by \eq{eq06}.

% Fig 2 a b 
\begin{figure}[h!]
\begin{center}
\includegraphics[width=0.750\columnwidth]{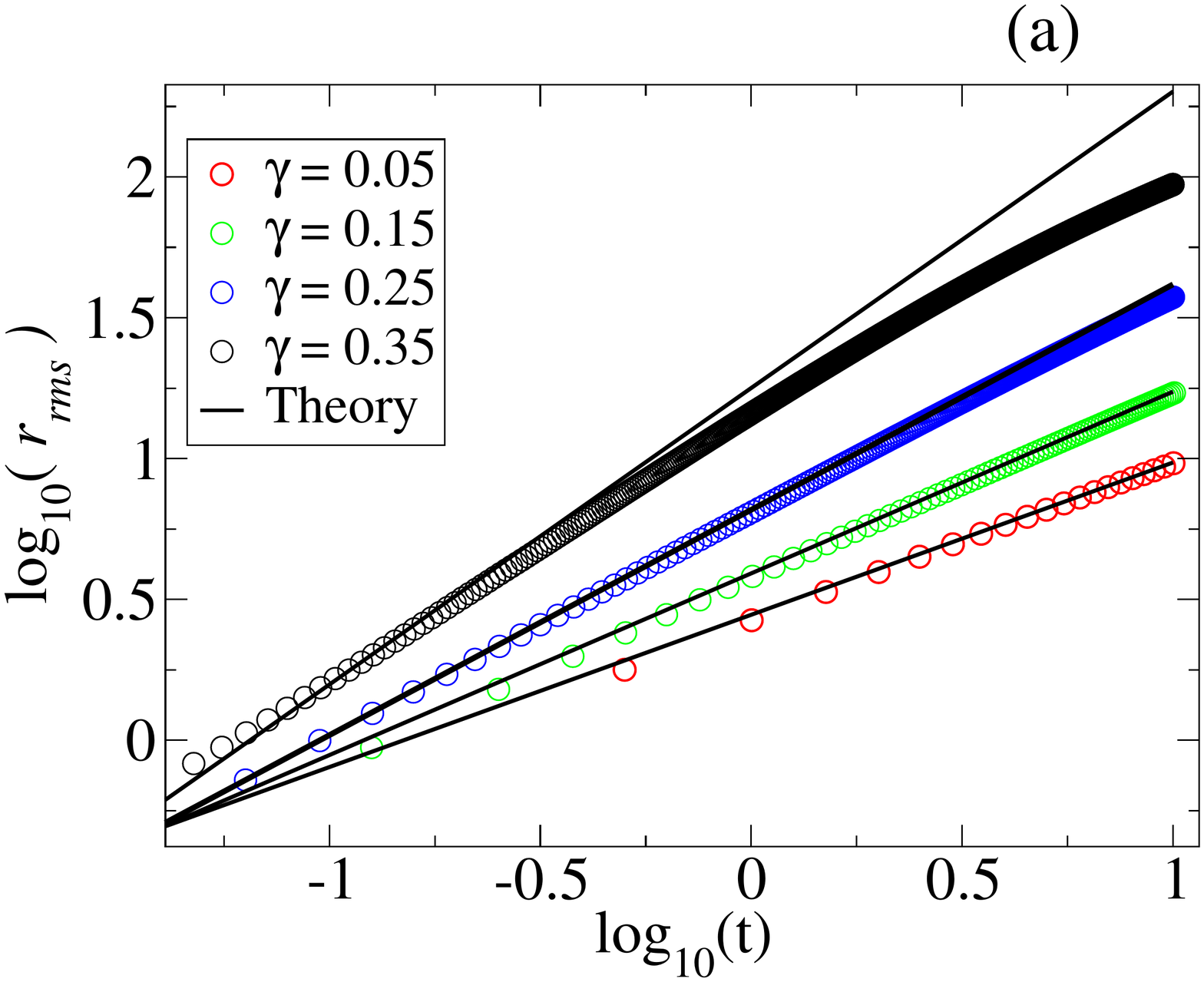}
\includegraphics[width=0.750\columnwidth]{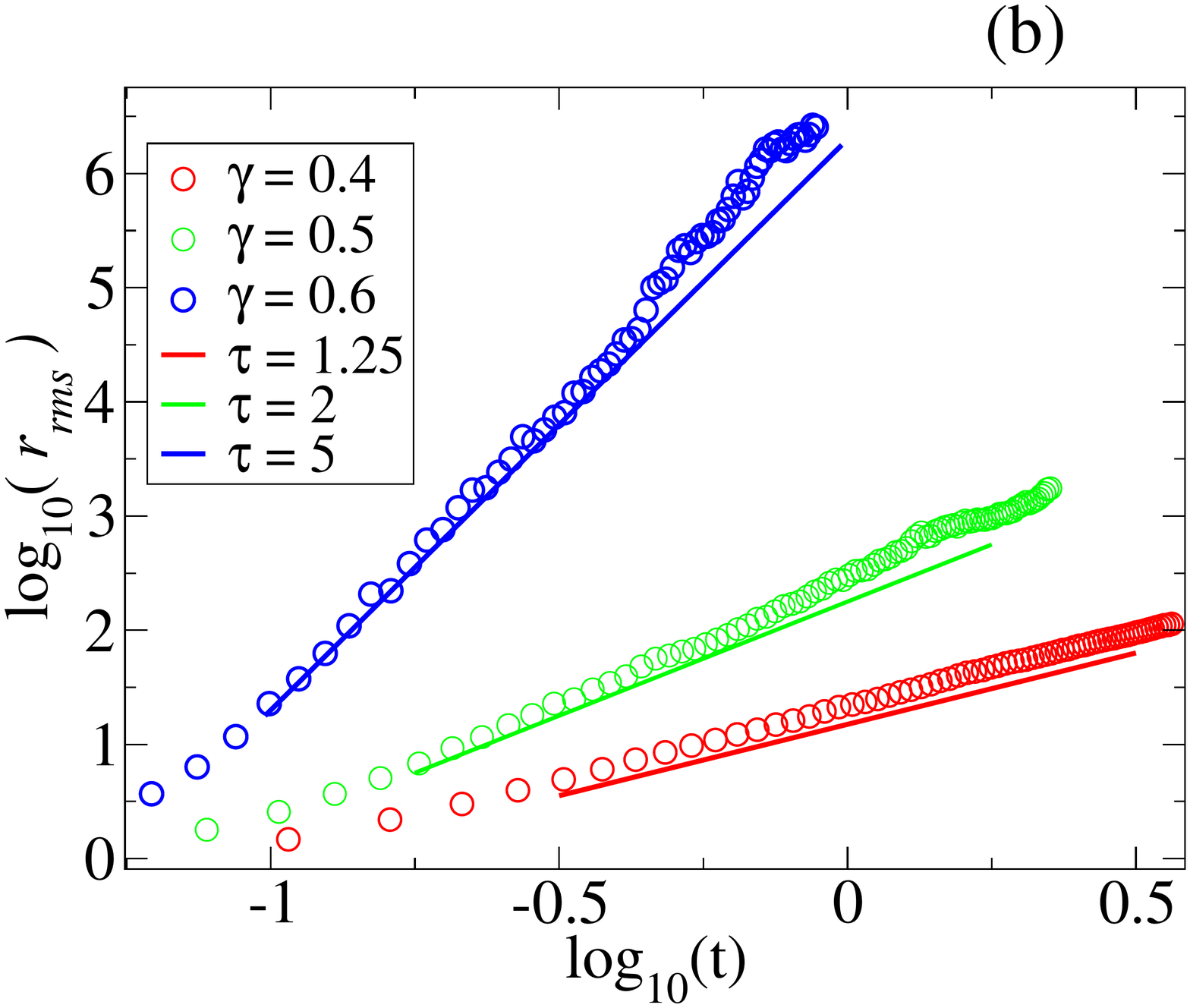}
\caption[]{ (a) Simulations of $r_{rms}$ compared to the 
  theoretical values of \eq{jmhvyd}   for the finite-interaction range model
  using $N_p=10^6$ particles   and $D_0=1$.
  (b) $r_{rms}$ resulting from the infinite-interaction
  range model using $N_p=1500$ particles. The solid lines show the predicted slope of \eq{tau}}
\label{figstep}
\end{center}
\end{figure}
 In the finite interaction range  model  $C$ is calculated by assuming a maximum interaction
range $\Delta x$ between particles. This is done  by
calculating $C$ onto a lattice with  lattice constant $\Delta x$:
The local value $C(\br_{n},t) $ at the discrete site $\br_{n}$ is
simply  $1/\Delta x^d$ times the number of
particles at positions $\bx_i$ that satisfy $|x_{i\alpha} -x_{n\alpha} |< \Delta
x/2$.
The  step length for a particle that is located at $\bx$ depends on the $C$-value at the
nearest lattice site.
The finite interaction range of this model has a
discretization effect: Once $C$ is so small that there is only  one-
or zero particles in each $\Delta x$-cell, the step length will always
be the same, and as a result, there will be a cross-over to normal
$r_{rms}\sim t^{1/2}$ diffusion, an effect that is observed in the
$\gamma = 0.35$ curve of Fig.~\ref{figstep} (a).

\begin{figure}[h!]
\begin{center}
\includegraphics[width=0.850\columnwidth]{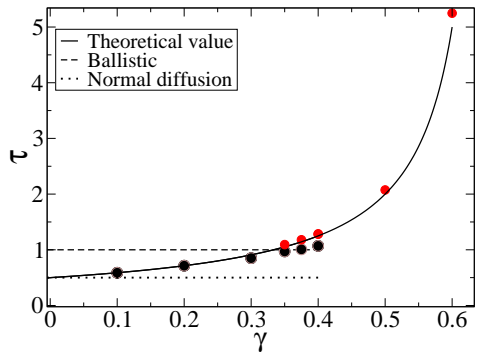}
\caption[]{   Simulation results 
for $\tau$ using the finite range $(\Delta x =1)$ model for $\gamma 
\le 0.6$  (black symbols), and the infinite-range model for 
$ \gamma =0.35 - 0.6$ (red symbols). The full line is the 
theoretical values. }
\label{figstigep}
\end{center}
\end{figure}
 The other, infinite interaction range model employs no lattice at all, but evaluates $C$ at
any particle position $\bx$ as
$ C(\bx ,t)= {N_r}/{V_r (\bx )} $
where $N_r \sim $10 is a fixed particle number and ${V_r (\bx )}$ is
the volume of the sphere that contains $N_r$  nearest
neighbors.  
There is no upper limit to the size of
${V_r (\bx )}$, and it is in this sense that the model has a potentially
infinite interaction range. When $\gamma \ne 0$ this model will never
cross
over to normal diffusive behavior. 

In Fig. \ref{jhga}  the analytic solution of \eq{eq11} is
plotted for  different $\gamma$-values.
The term 'explosive' seems an appropriate label for the behavior of 
the concentration for two reasons: First, as $\gamma \rightarrow 1/2$
close to the critical value of 2$/$3, the initial concentration $C(0,0)$
drops  by more than 10 
orders of magnitude in the same time that the negative $\gamma$
solutions (taken from Pattle \cite{pattle59}), drop by less than 2 orders. Second, the divergence of the 
integral in \eq{lkjiu} defining $r_{rms}(t)$
signals a cross-over to a regime where 
the break-away particles dominate the $r_{rms}(t)$-behavior at ever 
increasing step lengths.

In Fig. \ref{jhgb} the  data collapse anticipated in \eq{jhgi} is seen
to be satisfied. 
Figs. \ref{figstep} (a) and (b) demonstrate that the particle
displacement  is in fact characterized by  \eq{jmhvyd}, the difference
between Fig. \ref{figstep} (a) and (b),  being that the first figure
compares  simulations and the full analytic prediction of \eq{jmhvyd}, while the
hyperballistic transport shown in Fig. \ref{figstep}  (b), only
confirms the prediction of the $\tau$ exponent,  \eq{tau}.
Note that in Fig. \ref{figstep} (a)  the convergence to the prediction
of \eq{jmhvyd}, happens over a time that increases with $\gamma$,
signalling
the end of the regime where $r_{rms}(t)$ has an exact analytical expression.
 Figure  \ref{figstigep}  summarizes this comparison for the full range of
relevant $\gamma$-values, using the finite-range model for the
smaller- and the infinite range model for the larger $\gamma$-values. 

In conclusion, we have shown that particle interactions described
entirely in terms of their local concentration will yield
superdiffusion, and even hyperballistic diffusion when $d\ge 2$.

%--------------------------------------------------------------------
%--------------------------------------------------------------------
\begin{acknowledgements}
This work was partly supported by the 
Research Council of Norway through its Centers of Excellence funding 
scheme, project number 262644. 
\end{acknowledgements}
 \bibliographystyle{./../apsrev4}
 \bibliography{./../all,./refs}
\newpage 
\appendix
\section*{Supplemental material}
\begin{figure}[h!]
\begin{center}
\includegraphics[width=0.850\columnwidth]{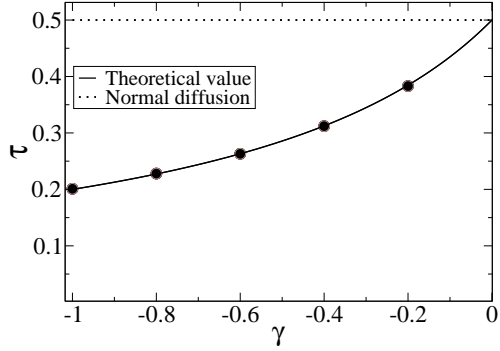}
\caption[]{   Simulation results 
for $\tau$ using the finite range $(\Delta x =1)$ model for $\gamma 
\le 0$ values (black symbols).The full line is the 
theoretical values. }
\label{fifiuggstigep}
\end{center}
\end{figure}
In Pattles classical 1959 paper \cite{pattle59} the $\gamma <0$ solution of \eq{eq01}
is not actually derived, but only written down.  So, for completeness
we derive it here along the same lines as those leading up  to
\eq{eq11}. 
  In the solutions thus derived 
$C(r,t)$ has a final support outside which it is strictly zero.
For any $\gamma$ the normalization
\begin{equation}
C_d
\int_0^{\infty}dr\,r^{d-1}
    C(r,t)=C_d     \int_0^\infty dy\, y^{d-1}p(y)  =N_p\; ,
\end{equation}
where we have used the isotropic nature of the problem to perform the 
angular integration and thus introduced  the geometric factor $C_d=1,2 \pi , 4\pi$ when $d=1,2,3$. 

 We see from equation (\ref{eq11}) that, for $\gamma<0$, the domain of the probability density, $p(y)$, is limited to $y<y_c=\sqrt{2k(\gamma -1)/\gamma}$, so that the normalization condition is
\begin{equation}
    \int_0^{y_c}dy\, y^{d-1}p(y)=\frac{N_p}{C_d}.
\end{equation}
yielding the normalization constant
\begin{equation}
    k=\left[N_p\left(\frac{\gamma }{2\pi(\gamma -1)}\right)^{\frac{d}{2}}\frac{\Gamma\left(\frac{d}{2}+1-\frac{1}{\gamma}\right)}{\Gamma\left(1-\frac{1}{\gamma}\right)}\right]^{\frac{2\gamma}{d\gamma-2}}\;,
\end{equation}
for $\gamma<0$.\\ 
We now combine results, using  Eqs.~(\ref{jhgi}), (\ref{kjgyr})  and (\ref{eq11}).
 to find the concentration field, $C(r,t)$
\begin{align}
    C(r,t)&=\Theta(r_c-r)\left(\frac{2-d\gamma}{1-\gamma}D_0C_0^{\gamma}t\right)^{-\frac{d}{2-d\gamma}}\nonumber\\
    &\left[k- \frac{\gamma}{2(\gamma -1)}\left(\frac{2-d\gamma}{1-\gamma}D_0C_0^\gamma t\right)^{-\frac{2}{2-d\gamma}}r^2\right]^{-\frac{1}{\gamma}},
\end{align}
where
\begin{equation}
    r_c=\left(\frac{2k(\gamma-1)}{\gamma}\right)^{\frac{1}{2}}\left(\frac{2-d\gamma}{1-\gamma}D_0C_0^{\gamma}t\right)^{\frac{1}{2-d\gamma}},
\end{equation}
By comparison, for $\gamma>0$ we have
\begin{align}
    C(r,t)&=\left(\frac{2-d\gamma}{1-\gamma}D_0C_0^\gamma t\right)^{-\frac{d}{2-\gamma}}\nonumber\\
    &\left[\frac{\gamma}{2(1-\gamma)}\left(\frac{2-d\gamma}{1-\gamma}D_0C_0^{\gamma}t\right)^{-\frac{2}{2-d\gamma}}r^2+k\right]^{-\frac{1}{\gamma}}, 
\end{align}
with $k$ given by \eq{jkhyre}  now.
Finally we find that for $\gamma<0$, $r^2_{rms} = A t^{2\tau}$ with 
\begin{align}
    A =\frac{\pi^{\frac{d}{2}}}{N_p}&k^{\frac{d}{2}+1-\frac{1}{\gamma}}
\frac{d}{2} 
                                       \frac{
\Gamma\left(1-\frac{1}{\gamma}\right)}{   
\Gamma\left(\frac{d}{2}+2-\frac{1}{\gamma}\right)}\nonumber\\
    &\left(\frac{2(\gamma-1)}{\gamma}\right)^{\frac{d}{2}+1}\left(\frac{2-d\gamma}{1-\gamma}D_0C_0^{\gamma}\right)^{\frac{2}{2-d\gamma}}.
\end{align}
\vspace{0.1cm}
In Fig. \ref{fifiuggstigep} this behavior is confirmed by simulations
using the finite-range model.

\end{document}